\documentclass[prb,twocolumn,superscriptaddress,floatfix,noshowpacs,10pt,longbibliography]{revtex4-1}%
\usepackage{graphicx,bm,times}
\usepackage{amsmath}
\usepackage{amsfonts}
\usepackage{amssymb}

\usepackage{bm}% bold math
\usepackage{color}
\usepackage{times}

\usepackage{xcolor}
\usepackage[%
  colorlinks=true,
  urlcolor=blue,
  linkcolor=blue,
  citecolor=blue
  ]{hyperref}

\begin{document}

\title{Ultra-high critical current densities, the vortex phase diagram and \\ the effect of granularity of the stoichiometric
high-$T_{\rm c}$ superconductor, CaKFe$_4$As$_4$ }

 \author{Shiv J. Singh}
 \affiliation{Clarendon Laboratory, Department of Physics,
 	University of Oxford, Parks Road, Oxford OX1 3PU, UK}

 \author{Matthew Bristow}
 \affiliation{Clarendon Laboratory, Department of Physics,
 	University of Oxford, Parks Road, Oxford OX1 3PU, UK}

 \author{William Meier}
 \affiliation{Ames Laboratory, Iowa State University, Ames, Iowa 50011, USA}
 \affiliation{Department of Physics and Astronomy, Iowa State University, Ames, Iowa 50011, USA}

  \author{Patrick Taylor}
 \affiliation{Clarendon Laboratory, Department of Physics,
 	University of Oxford, Parks Road, Oxford OX1 3PU, UK}

   \author{Stephen J. Blundell}
 \affiliation{Clarendon Laboratory, Department of Physics,
 	University of Oxford, Parks Road, Oxford OX1 3PU, UK}

 \author{Paul C. Canfield}
 \affiliation{Ames Laboratory, Iowa State University, Ames, Iowa 50011, USA}
 \affiliation{Department of Physics and Astronomy, Iowa State University, Ames, Iowa 50011, USA}

 \author{Amalia I. Coldea}
 \affiliation{Clarendon Laboratory, Department of Physics,
 	University of Oxford, Parks Road, Oxford OX1 3PU, UK}

\begin{abstract}
	We present a comprehensive study of the critical current densities and
the superconducting vortex phase diagram in the stoichiometric superconductor CaKFe$_4$As$_4$
which has a critical temperature of $\sim\,$35 K. We performed detailed magnetization measurements
both of high quality single crystals for different orientations in an applied magnetic field  up to 16~T
and for a powder sample.
We find an extremely large  critical current density, $J_{c}$, up to $10^{8}\,$A/$\mathrm{cm}^{2}$ for single crystals when $H||(ab)$ at 5$\,$K,
which remains robust in fields up to 16$\,$T,  being the largest of any other iron-based superconductor.
 The critical current density is reduced by a factor 10 in single crystals when $H||c$ at 5~K
 and significantly suppressed by the  presence of grain boundaries in the powder sample.
We also observe the presence of the fishtail effect in the magnetic hysteresis loops of single crystals when $H||c$.
The flux pinning force density and the pinning parameters suggest
that the large critical current could be linked to the existence of point core  and surface pinning.
Based on the vortex phase diagram and the large critical current densities, CaKFe$_4$As$_4$  is now established
 as a potential iron-based superconductor candidate for practical applications.
\end{abstract}
\date{\today}
\maketitle

\section{Introduction}
The discovery of iron based superconductors \cite{{Kamihara2008}, {Hosono2015c}} has led
to important advances in superconductivity not only because of their unconventional superconducting mechanism with
rather high superconducting transition temperatures, but also due to their potential
for high magnetic field applications, such as superconducting wires and tapes.
  The bulk iron-based superconductors have a high transition temperature of up to 58~K \cite{{Zhi-An2008a}, {Singh2013a}}
  and very high upper critical field ($H_{c2}$),  up to 100$\,$T \cite{{Jaroszynski2008c}, {Jaroszynski2008b}}.
Furthermore, their high upper critical field, small anisotropy, good grain boundary connectivity and high critical current density under high magnetic field
are all important factors contributing to their suitability for practical applications \cite{{Ma2012a}, {Shimoyama2014a}}.

 Among iron-based superconductors, the recently discovered CaKFe$_4$As$_4$
  is a member of a new
 1144 structural family of iron-based superconductors, being a stoichiometric optimally-doped superconductor,
 with a transition temperature, $T_{c}$, around 35~K and large upper critical field up to 90$\,$T \cite{Iyo2016,Meier2016}.
A number of transport, thermal and thermodynamics experiments on
CaKFe$_4$As$_4$ do not show any structural transition or magnetic order  \cite{Iyo2016,Meier2016}.
 This system displays similar superconducting properties to those of the optimally-doped
 Ba$_{1-x}$K$_{x}$Fe$_2$As$_2$
(BaK122) system \cite{Rotter2008b}.
CaKFe$_4$As$_4$ has a tetragonal structure ($P4/mmm)$,
where Ca and K layers stack alternatively across the $\mathrm{Fe}_2$$\mathrm{As}_2$ layer along the $c$ axis \cite{Iyo2016,Meier2016};
Fe-As bond distances are different for the As atoms above and below the Fe plane, in contrast to the 122 system
that has equivalent Fe-As distances. The lowering in symmetry has significant consequences on the electronic structure
of CaKFe$_4$As$_4$, being that of a compensated metal with a large number of electronic bands close to a nesting instability \cite{Mou2016}
as well as on its magnetic structure which forms an unusual spin-vortex crystal order by Ni or Co doping  \cite{Meier2018}.

 Generally, iron-based superconductors possess a layered structure similar to that of the high-$T_{c}$ cuprates.
 Superconductivity  in these systems often appears out of an antiferromagnetic state with bad metallic behaviour, found in the parent compounds,
 which is  suppressed by electron (or hole) doping  \cite{Hosono2015c,Johnston2010b,Hirschfeld2016b} or by application of external pressure \cite{Hosono2015c},
 giving rise to complex superconducting phase diagrams \cite{Hosono2015c,Johnston2010b,Hirschfeld2016b}.
 The understanding of the superconducting mechanism of unconventional superconductors
  is challenging due to different competing electronic phases,
   but the existence of few stoichiometric iron-based superconductors provides a unique opportunity
   to study this phenomenon in the absence of additional extrinsic effects introduced by chemical substitutions.

The critical current density, $J_{c}$,  is an essential property of the high-$T_{c}$ superconductors in order to identify
the best  candidates for practical applications.
 A high $J_{c}$ exceeding $10^{5}\,$A/$\mathrm{cm}^{2}$ even under high magnetic field above 10~T is found in many optimally-doped
  iron-based superconductors \cite{Ma2012a}. The high transition temperatures, large magnetic penetration depth, vortex motion and thermal fluctuation generally plays an important role in the critical current properties, similar to the high-$T_{c}$ cuprates. These phenomena cause interesting effects in the vortex dynamics, such as giant-flux creep, thermally activated flux flow and  the presence of a second magnetization peak or the fish-tail effect \cite{Ishida2017,Blatter1994a}.
  The ability to carry the critical current is governed by the vortex pinning strength and motion.
  Different kinds of pinning determine the elastic or plastic motion of vortex lattice,
  formation of vortex glass, vortex melting and order-disorder phase transitions   \cite{Blatter1994a,Feigelman1989}.
As vortex motion is strongly affected by the effects of disorders in superconductors, it is important to understand what happens
in a stoichiometric and clean high-$T_{\rm c}$ superconductor, such as CaKFe$_4$As$_4$.

In the present study, we report the superconducting properties and the critical current densities of  CaKFe$_4$As$_4$  in both single crystals and polycrystalline samples
 using magnetization measurements in magnetic fields up to 16$\,$T. We find that CaKFe$_4$As$_4$ exhibits an ultra-high critical current density close to $J_{c}$$\sim$$10^{8}\,$A/$\mathrm{cm}^{2}$ at 5$\,$K, when the magnetic field is parallel with the conducting plane,
 as extracted from isothermal magnetic hysteresis loops.
 Critical current density is reduced by a factor 10 in single crystals,  when the magnetic field is parallel with the $c$-axis at 5$\,$K,
  and it is significantly suppressed by the effect of grain boundaries in powder samples. Our study compares the behaviour of the lower critical field ($H_{c1}$) and the upper critical field ($H_{c2}$) of single crystals
 and the vortex phase diagram of single and polycrystalline samples.
 We detect the presence of a second magnetization peak  in the magnetic hysteresis loops of single crystals when $H||c$.
The flux pinning force density and the pinning parameters suggest the existence of surface and
point core pinning of
vortices.
Our results suggest that CaKFe$_4$As$_4$, with such large critical current densities
and very high upper critical fields, is an exciting new candidate  for practical applications.

\section{Experimental details}

The single crystal CaKFe$_4$As$_4$ samples were grown by the FeAs self-flux method and polycrystalline sample were synthesized by solid state reaction method,
as detailed elsewhere \cite{Meier2017,Meier2016,Iyo2016}. The single crystals have rectangular shapes for both magnetic and resistivity measurements, whereas the
polycrystalline sample was pressed powder in a cylindrical shape. Magnetic measurements were carried out using the VSM (Vibrating Sample Magnetometer) option of a Physical Property Measurement System (PPMS-Quantum Design) with the magnetic fields up to 16$\,$T.
The magnetic isotherms were recorded up to 16~T with ramping rate of 3~mT/s, at several temperatures ranging from 2 up to 40~K. All the hysteresis loops at constant temperatures, $M$($H$), and temperature dependent magnetization in constant magnetic field, $M$($T$), were performed after cooling the sample in zero magnetic field from above $T_{c}$.

Magnetization hysteresis loops allow us to extract the critical current density, $J_{c}$, as a function of magnetic field
using the extended Bean's critical state model \cite{Bean1985}.
The vortex pinning force density can be estimated using $F_{P}$ = $J_{\rm c} \times \mu_{0} {H} $  \cite{Dew-Hughes1974}.
For a rectangular crystal, the magnetic field dependence of the critical current density, $J_c$ (in units of A/$\mathrm{cm}^{2}$)
is given by the relation  $J_c = 2 \Delta M$/[$a$(1-$a/3b$)] (for \textit{a}$<$\textit{b});
$\Delta M=M_{up} - M_{dn}$, and $M_{up}$ and $M_{dn}$ are the magnetization while magnetic
fields are decreasing and increasing, respectively.
The magnetization $\Delta$\textit{M} expressed in A/cm
is obtained from
the magnetic moment divided by the sample volume, $V=a \times b \times c$,
 where $a$ and $b$ are in-plane sample dimensions perpendicular to the applied
magnetic field and $c$ is  sample dimension, parallel to the applied magnetic field.
We have measured a single crystal of CaKFe$_4$As$_4$
with the following  dimensions: $a= 1.09$~mm, $b=1.365$~mm and $c= 0.0025$~mm, measured with the magnetic field aligned along $c$-axis  ($H||c$)
and  \textit{a} = 0.0025~mm, \textit{b} = 1.09~mm and  \textit{c} = 1.365~mm, measured with the magnetic field aligned in the $ab$-plane ($H||$($ab$)).
The sample thickness of the crystal was evaluated using Gaussian fits to the different optical image measurements across the sample, as shown in Fig.\ref{fig1SM}.
For the cylindrical pressed powder sample in order to estimate the critical current, $J_c$,
we used the expression  $J_c$ = 3 $\Delta$\textit{M}/\textit{d},
where $d$ = 5.03~mm and $l$ = 4.08~mm are the diameter and length of the cylinder, respectively \cite{Bean1985}.
 The average grain size for the powder sample is around 1.7(7)~$\mu$m.

\begin{figure}[htbp]
	\centering
	\includegraphics[trim={0cm 0cm 25cm 0cm}, width=0.85\linewidth]{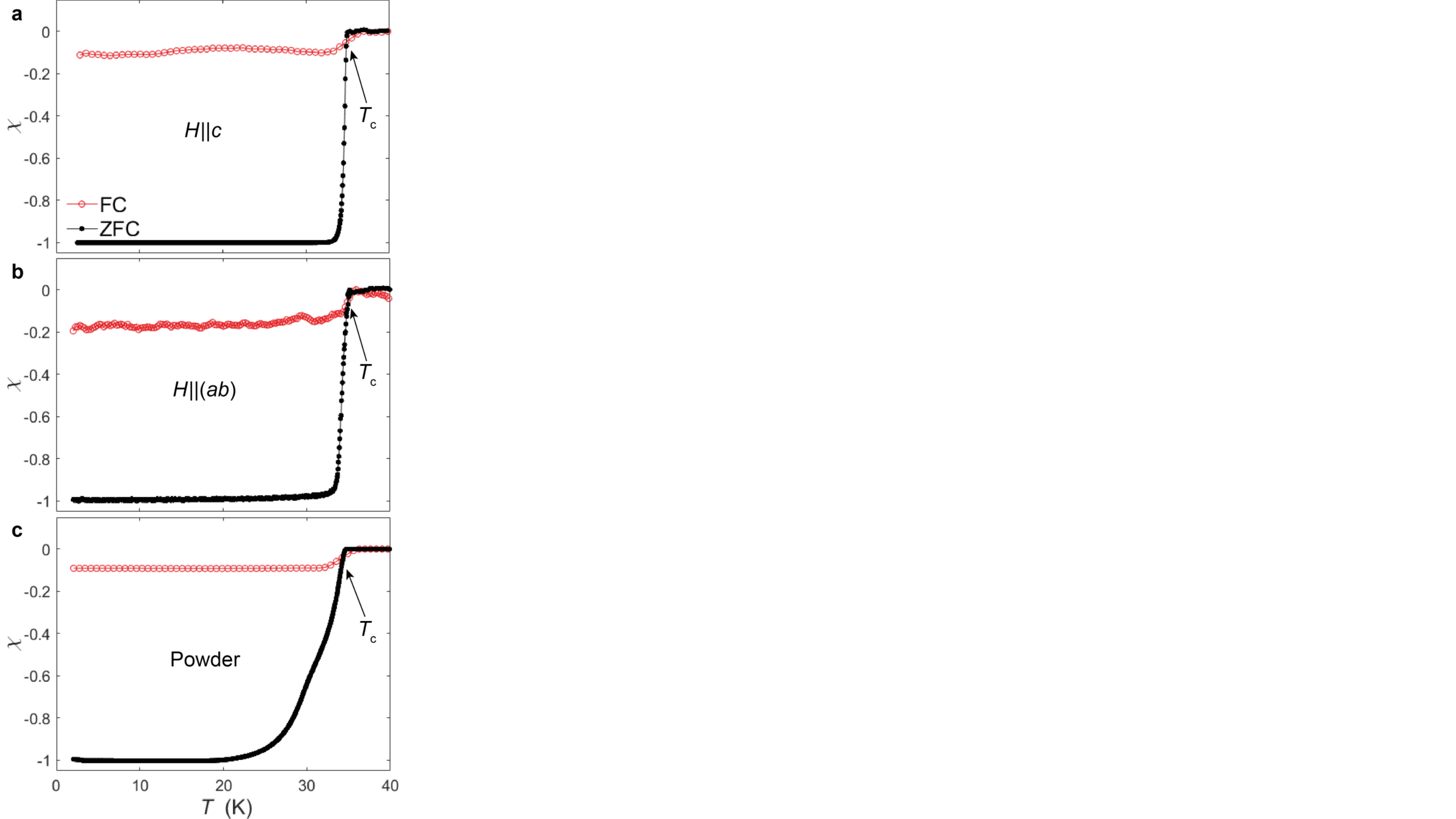}
	\caption{Temperature dependence of magnetic susceptibility ($\chi$) deduced from zero-field cooled  and field-cooled  magnetization
		measurements for  single crystal when  (a) $H || c$ and (b) $H||$($ab$) in 5~mT	as well as (c) the powder sample in 2~mT.
		The arrows indicate the positions of the onset values of the critical temperature, $T_{\rm c}$.}
	\label{fig1}
\end{figure}

To estimate the lower critical field, $H_{c1}$, $M$ ($H$) loops were measured up to 0.5$\,$T  sweeping the field with a slow rate of 0.2$\,$mT/s at each temperature. Before each measurement the sample was warmed up above $T_{c}$, and the remnant field of the magnet is reduced before cooling down to the target temperature. The susceptibility data, $\chi=M/H$ has been corrected for the demagnetization effect \cite{Chen2002}.  For an applied external field $H_a$, the internal field
$H_i$ = $H_a - N_{\rm eff} M$, where $N_{\rm eff}$ is the effective demagnetization factor.
For the accurate determination of $H_{c1}$, we subtracted the value of magnetization
 obtained by the low-field magnetization slope from the magnetization (\textit{M}) for each isotherm \cite{Naito1990}.
The deviation point of
 of magnetization versus field from linear behaviour gives the value of the first penetration field, $H_{c1}^*$.
 We estimate the effective demagnetization factor for our single
 crystal to be $N_{\rm eff}$ = 0.995 when $H||c$,  which is similar to the calculated demagnetization factor for this orientation \cite{Chen2002}.
 When  $H||$($ab$)  the value of $N_{\rm eff}$ = 0.70
 which is extracted from zero-field cooled data assuming that in the orientation the sample is a
 perfect diamagnet as for $H||c$.
  $N_{\rm eff}$ = 0.41 for the cylindrical polycrystalline sample is extracted both from
calculations and experiment \cite{Chen2002,Brandt1999}.
 The upper critical field, $H_{c2}$,  is extracted both from the temperature dependence of magnetization
 at different constant magnetic field up to 16~T.

\section{Results and discussion}

The temperature dependence of the magnetic susceptibility ($\chi$) from zero-field-cooled (ZFC) and field-cooled (FC) magnetization measurements in low magnetic field (2 or 5 mT) for single crystal (when $H||c$ and $H||$($ab$)) and for the powder sample of CaKFe$_4$As$_4$ are shown in Fig. 1(a)-(c).
A perfect diamagnetic state is reached at low temperatures for all measured samples.
The critical temperature, $T_{c}$, is defined around 35~K
from the bifurcation point between ZFC and FC branches of the magnetization in Fig.~\ref{fig1}.
The width of the superconducting transition is very narrow for the single crystal
($\Delta T_c  \sim $1~K) suggesting a high quality of this sample,
as shown by the X-ray and resistivity data in the Supplemental Material in Fig.\ref{fig1SM}.
The powder sample also show perfect diamagnetism
at low temperatures but it has  a much broader transition of ($\Delta T_c  \sim $12~K), likely due to pinning to grain boundaries and
grain size effects, as seen in other iron-based superconductors \cite{Singh2013}.
No secondary phases were identified in our powder sample, as shown in  Fig.~\ref{fig2SM}.
The small diamagnetic signal in the FC data reflects the strong pinning nature of the sample and permanent fluxed trapped inside it.

\begin{figure*}[htbp]
	\centering
\includegraphics[trim={0cm 0cm 20cm 0cm},width=0.7\linewidth,clip=true]{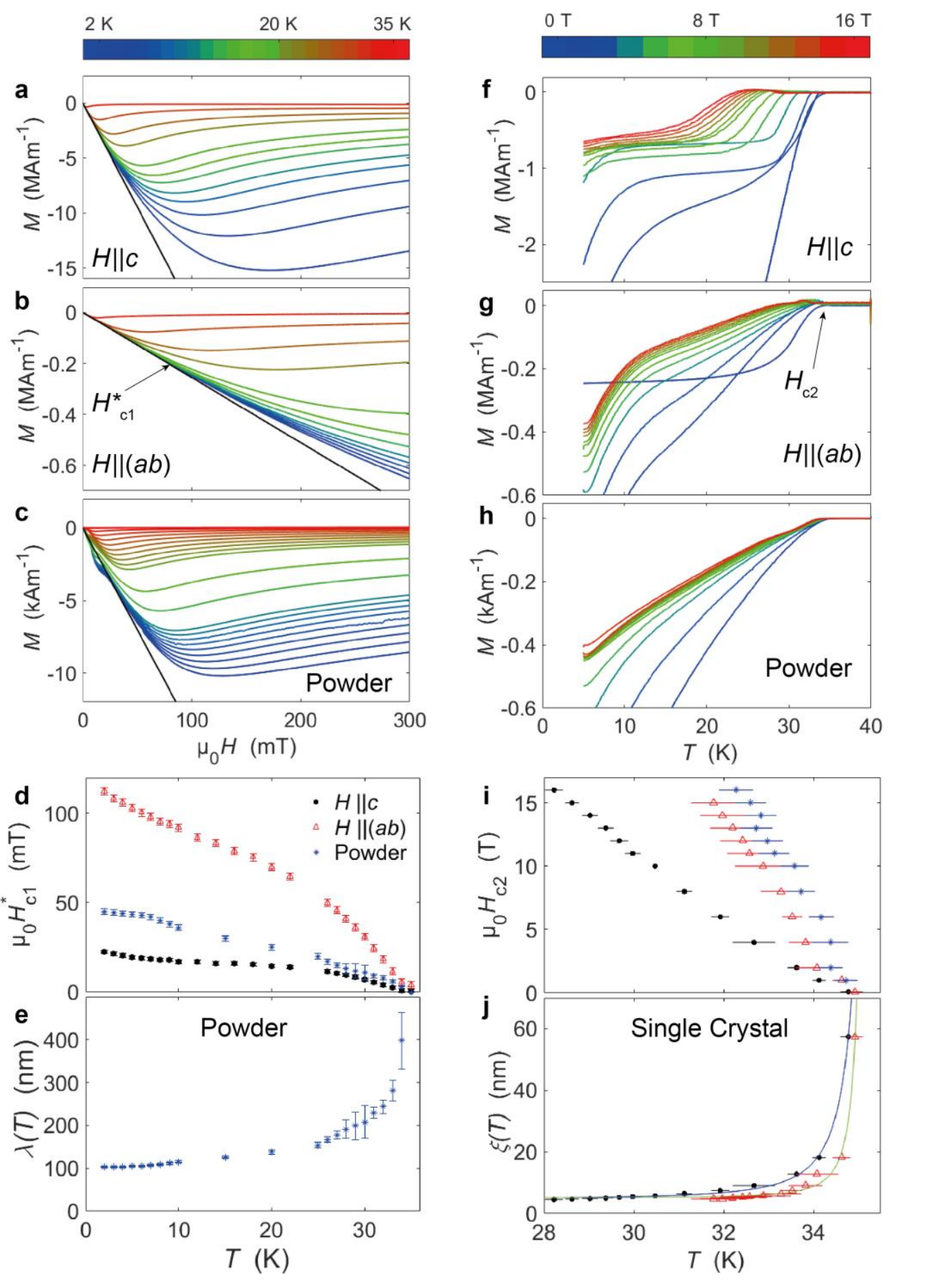}
	\caption{Magnetization, $M$, versus magnetic field at constant temperatures for the single crystal when (a) $H || c$ and (b) $H||$($ab$) as well as for (c) the powder sample of CaKFe$_4$As$_4$.
The arrow in (b) shows the point of deviation of magnetization from linear dependence in the Meissner state,  defined as the first penetration field $H_{\rm c1}^*$.
(d) The temperature dependence of the lower critical field $H_{\rm c1}^*$ measured for the single crystal
in two orientations in magnetic field as well as the powder sample.
(e) The penetration depth, $\lambda$, for the powder sample.
 The temperature dependence of magnetization
in different magnetic fields up to 16~T for the
single crystal in two orientations in (f)-(g)
and for the powder sample in (h).
(i) The upper critical field $H_{c2}$ versus temperature for the single crystal and  the powder sample. %with 5\% error $\textbf{(d)}$.
(j) The coherence lengths, $\xi_{ab}$ and $\xi_{c}$,  for a single crystal of CaKFe$_4$As$_4$. }
 	\label{fig2}
\end{figure*}

The temperature dependence of the superconducting lower critical field, $H_{\rm c1}$, is determined using low-field magnetic hysteresis measurements, as shown in Fig.~\ref{fig2}(a)-(c). We measured the field dependence of the magnetization in the superconducting state at different temperatures, with the external field in the ($ab$)-plane and along the $c$-axis and also for the powder sample.
The linear variation of the magnetization is a signature of the Meissner state up to $H_{\rm c1}$,
 above which the vortices starts to form. The temperature dependence of the extracted $H_{\rm c1}$ for the single crystal and the powder sample
       is shown in Fig.~\ref{fig2}(d). The values of $H_{\rm c1}$ for $H||$($ab$) is  higher than that for $H || c$.
       The $H_{c1}$ value for polycrystalline sample lies in between these two orientation of the single crystal (as shown in Fig.~\ref{fig2}d),
       as expected from the powder averaging effects,
       and it was extracted by fitting a linear dependence to the magnetization above the anomaly caused by the grain boundaries.
   The $H_{c1}$($T$) powder data show a very weak temperature dependence at low temperatures below 10~K, consistent
   with recent $\mu$SR results \cite{Biswas2017}.

\begin{figure*}[htbp]
	\centering
	\includegraphics[trim={0cm 0cm 10cm 0cm},width=0.9\linewidth,clip=true]{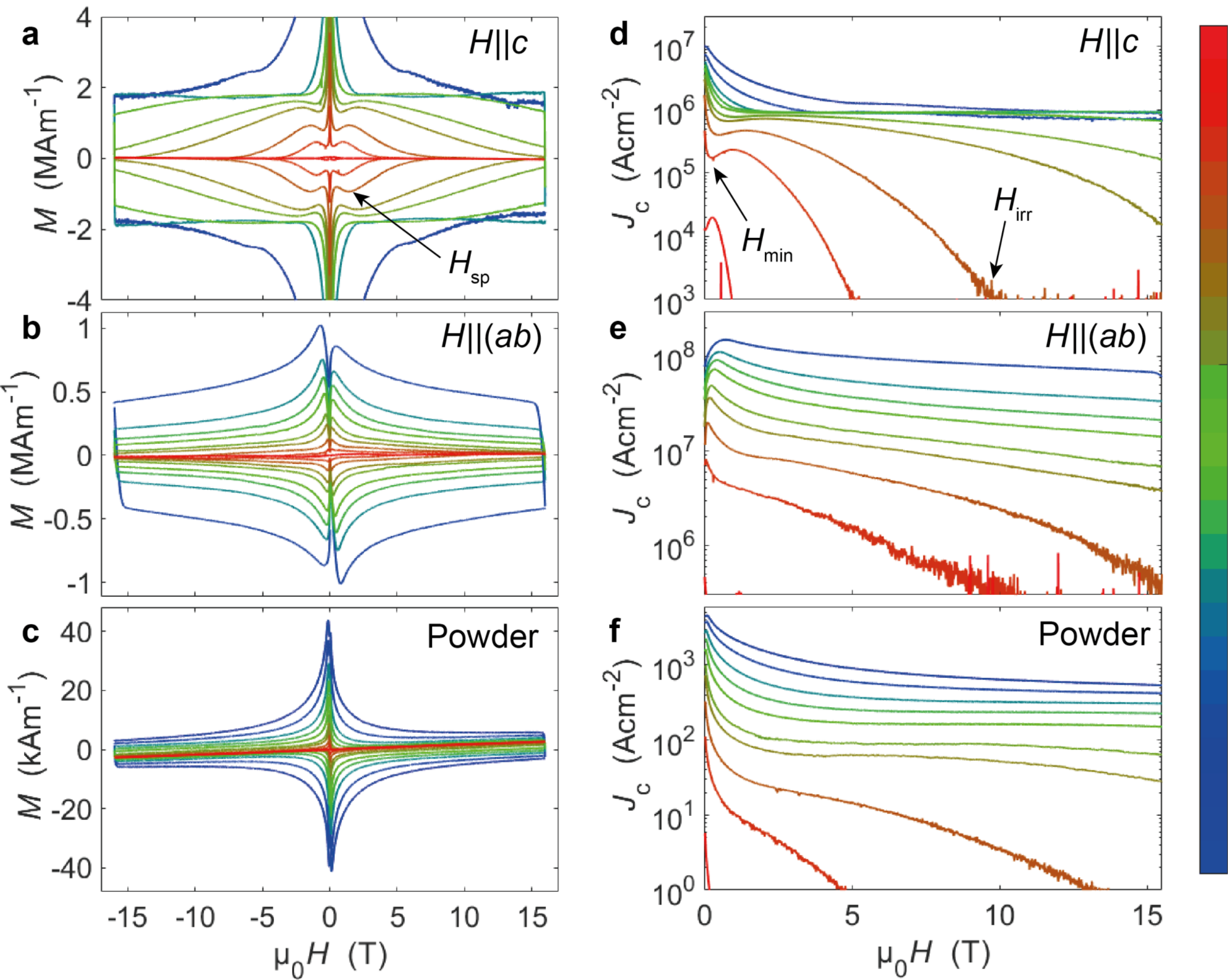}
		\caption{The isothermal magnetization loops, $M$($H$), for
		the single crystal in two orientations with (a) $H || c$ and (b) $H||$($ab$) and
 (c) the powder sample of CaKFe$_4$As$_4$.
The critical current density, $J_{c}$, calculated from $M$($H$) using the Bean model \cite{Bean1985}, plotted as a function of magnetic field at constant  temperatures for the single crystal (d) when $H || c$	and (e) $H||$($ab$)	 and (f)  the powder sample, respectively.
		The arrows indicate the positions of $H_{\rm sp}$ in (a) and $H_{\rm irr}$ and
		$H_{\rm min}$ in (d).}
	\label{fig3}
\end{figure*}

To determine the upper critical field, $H_{\rm c2}$, we have measured zero-field cooled (ZFC) magnetization in the
temperature range between 2 to 40~K for various values of the magnetic field up to 16~T,
shown in Figs.~\ref{fig2}(f)-(h) for the single crystal and the powder sample, respectively.
The upper critical field, $H_{\rm c2}$($T$),  as a function of temperature near $T_{\rm c}$
is defined by the value at which magnetization starts to deviate from the high-temperature paramagnetic value.
Just below the critical temperature, we noticed a small jump in the temperature dependence
of zero-field magnetization for both the single crystal and the powder sample,
being  more pronounced as the field increases. This effect may be caused by the flux jumps.

 Fig.~\ref{fig2}(i) shows the $H-T$ phase diagram of CaKFe$_4$As$_4$ calculated from magnetization data.
Since the transition temperature does not shift much with magnetic field
 towards low temperatures, it indicates a very high value of $H_{\rm c2}$(0).
To estimate $H_{\rm c2}$(0) we use the Werthamer-Helfand-Hohenberg (WHH) formula \cite{Effects1963}
close to $T_c$, $\mu_0 H_{\rm c2}$(0) = -0.69$T_c$(d$ \mu_0 H_{c2}$/d\textit{T}).
The slope of upper critical field (-d$\mu_0 H_{\rm c2}$/d\textit{T}) (see Table~SM1)
ranges between 2.6(1) for $H || c$,
to 5.7(7) for $H||$($ab$) for the single crystal and
10.1(5) T$K^{-1}$ for the powder sample.
Using the transition temperature as $T_c$ = 35~K,
we find $ \mu_0 H_{\rm c2}$ at zero temperature of $\sim 62$~T
when $H || c$ and a much larger value of 135~T  when $H||$($ab$) .
The powder sample show a significantly larger
upper critical field of 241~T, a factor of $\sim 4$ larger than in single crystals.
We have also measured the resistivity for a different single crystal,
as shown in Fig. 1SM(c).
The slope of upper critical field from transport data measured in magnetic field (not shown here \cite{Bristow2018})
gives much larger upper critical fields,
$ \mu_0 H_{c2}$(0) = 82~T, and 168~T for $H || c$ and $H||$($ab$), respectively,
similar to those reported previously \cite{Meier2016}.

 Using the zero temperature value of upper critical field $H_{c2}$(0), the corresponding value of $\mu_{0}$$H_{c2}$(0)/$k_B$$T_c$  is close to 1.8
 for $H || c$,  and increases to 3.9 when $H||$($ab$) and 6.9 T$K^{-1}$ for the powder sample.
The Pauli limit is reached when the condensation energy
is overcome by the Zeeman energy for
normal  electrons, defined as $\mu_{0}H_{p}$/$ T_c$ = 1.84 T K$^{-1}$ in the case of singlet pairing and weak spin-orbit coupling \cite{Wang2008b,Clogston1962}.
 In order to estimate the superconducting order parameter we used the Ginzburg-Landau (GL) approach.
 This gives a coherence length
 extrapolated to zero temperature
 as   $\xi_{\rm ab}$=1.9 nm, $\xi_{\rm c}$=0.87 nm for the single crystal ( as shown in Fig.~\ref{fig2}j)
    and $\xi=0.97$ nm for the powder sample.
  The extracted mean free path is
$\ell= 15.7$ nm (using carrier concentration reported in Ref.~\onlinecite{Meier2016})
based on the extrapolated normal resistivity at zero temperature
of $\rho_n \sim 15.9$~$\mu \Omega$cm (see Fig.~SM1(c)).
This suggests that   our crystals lie within the clean limit with $\xi$(0)$ \ll  \ell$.

The $H_{\rm c1}$ values extracted in Fig.~\ref{fig2}(d) were used to calculate the London penetration depth as a function of temperature, $\lambda(T)$.
In order to estimate the penetration depth for the powder sample, we correct the value for demagnetizing effects (as detailed earlier using $N_{\rm eff}$=0.41)
 and estimate the value of the Ginzburg-Landau parameter as $\kappa \sim 88$.
This gives a low temperature London penetration depth
$\lambda$(0) of 104~nm ( as shown in Fig.~\ref{fig2}e), which is similar
to the value of 133~nm, extracted from London penetration depth studies in single crystals
\cite{Cho2017}. Muon relaxation studies
estimate the penetration depth as 209~nm for a single crystal \cite{Khasanov2018} and
286~nm for a powder sample \cite{Biswas2017}.

\begin{figure*}[htbp]
	\centering
    \includegraphics[trim={0cm 0cm 0cm 0cm},width=1\linewidth,clip=true]{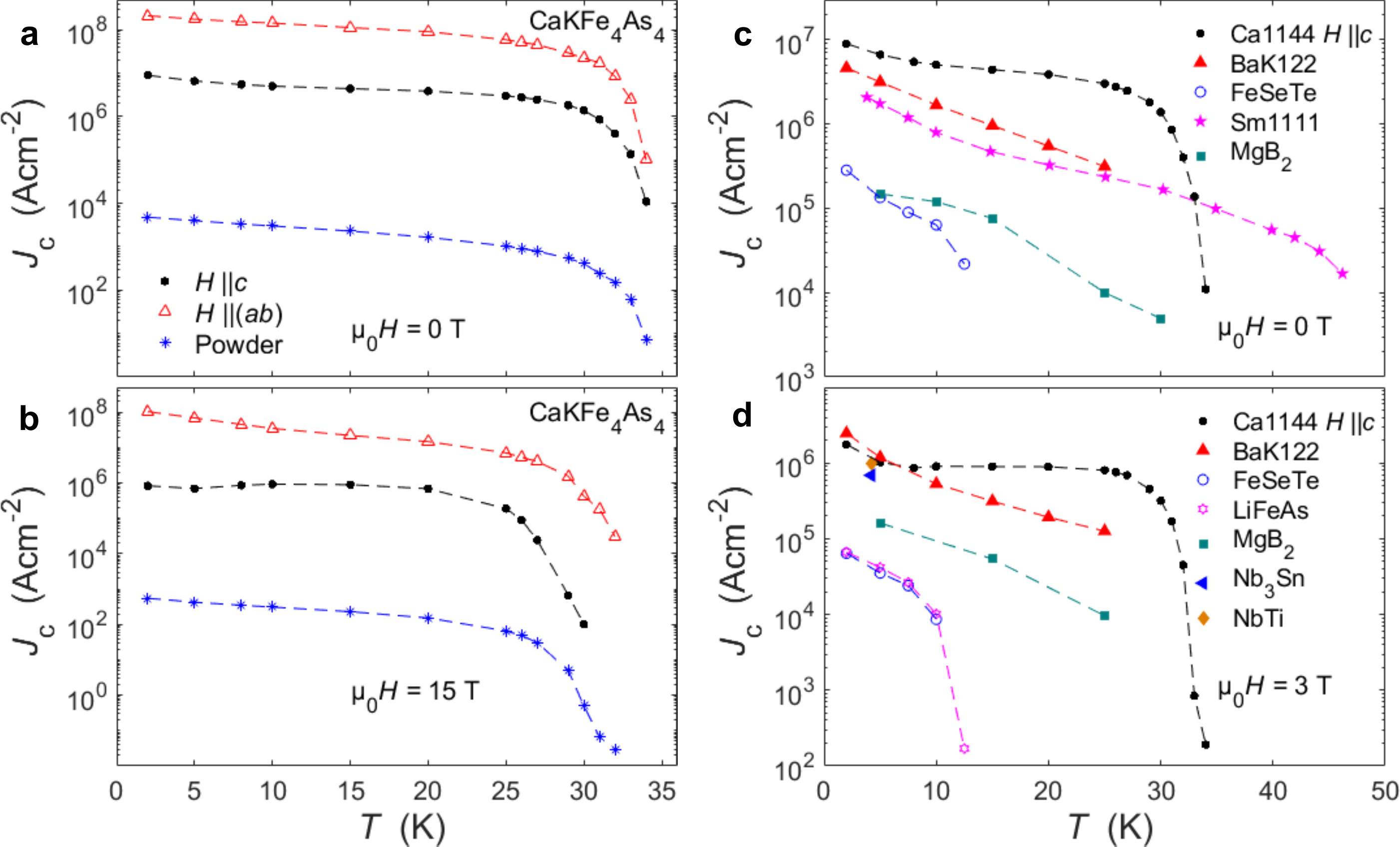}
		\caption{The temperature dependence of critical current density, $J_c$, extrapolated at (a) $\mu_0 H$ = 0 T and (b) $\mu_0 H$= 15 T
of  the single crystal in the two orientations in magnetic field  and of the powder sample of CaKFe$_4$As$_4$.
Comparison between the temperature dependence of the critical current density $J_c$ (calculated using the Bean model \cite{Bean1985}) in single crystals of CaKFe$_4$As$_4$ with those reported for different iron-based superconductors (SmFeAs(O,F)\cite{Fang2013}, (Ba,K)Fe$_2$As$_2$ \cite{Yang2008}, Fe(Se,Te) \cite{Taen2009}, LiFeAs\cite{Pramanik2011a}) and conventional
superconductors (MgB$_2$ \cite{Eisterer2007,Moore2003,Buzea2001}, Nb$_3$Sn \cite{Buzea2001}, NbTi \cite{Heussner1997})
in magnetic field of (c) 0~T and (d) 3~T for $H || c$.}
	\label{fig4}
\end{figure*}

Figs.~\ref{fig3}(a)-(c) show the $M$($H$) loops at constant temperatures between 2 to 35~K in a magnetic field up to 16~T
for the single crystal in the two orientations
($H||$($ab$) and $H || c$) and for the powder sample, respectively.
The hysteresis areas decreased with increasing temperature and the symmetrical shape of $M$($H$) loops  imply the existence of flux
pinning centres, suggesting that the magnetization is dominated by bulk pinning rather than surface and geometrical barriers \cite{Dew-Hughes1974}.
Furthermore, the magnetic hysteresis loop in Figs.~\ref{fig3}(b) when $H||$($ab$) displays a dip near zero magnetic field which may be caused by the highly inhomogeneous field distribution in the vortex state \cite{Tamegai2012} and the anisotropy of $J_c$ \cite{Mikitik2005}.

The second magnetization peak (SMP) can be easily seen for the single crystal below 26~K when $H || c$, as shown in Fig.~\ref{fig3}(a).
With decreasing temperature, the peak moves to higher magnetic fields beyond the maximum applied magnetic field of 16~T.
In the region between the valley and the peak, the width of the irreversible magnetization, $ \Delta M$, in
Fig.~\ref{fig3}(a) extends with increasing magnetic field and shows a clear fishtail effect.
This  effect was observed in other superconducting single crystals, such as cuprates \cite{Welp1990,Yeshurun1994}, Nb$_3$Sn \cite{Lortz2007}, MgB$_2$ \cite{Pissas2002}
and iron-based superconductors, such as BaK122 \cite{Yang2008}  and LiFeAs \cite{Pramanik2011a}.
Although its origin is not fully explained, the fishtail effect is strongly dependent on the sample orientation in externally applied magnetic field, being diminished when $H||$($ab$). Importantly, the width of the $M$($H$) loops for $H||$($ab$) (Fig.~\ref{fig3}(b)) is much larger than of the loops for $H || c$ (Fig.~\ref{fig3}(a)) or
the powder sample (Fig.~\ref{fig3}(c)), implying that the pinning mechanisms is anisotropic and strongly affected by the granularity.
Similar effects caused by anisotropy of flux pinning were observed in other iron-based superconductors, such as Fe(Se,Te) \cite{Taen2009} and SmFeAs(O,F) \cite{Fang2013}.

According to the extended Bean model \cite{Bean1985}, the critical current density ($J_c$) can be determined from the hysteresis loop width ($\Delta$\textit{M}).
 As the critical current remains rather constant with the change of magnetic field for the low
temperature magnetization isotherms, the extended Bean's model  gives a good estimation of the critical current density
in our case. Other models that take into account the field variation of the critical current density and
the vortex dynamics were also developed by Kim in Ref.~\onlinecite{Kim1963} and Kim-Anderson and in Ref.~\onlinecite{Anderson1964}.
Figs.~3(d)-(f)  show the field dependence of $J_c$ at constant temperatures obtained for the single crystal  for
different orientations in magnetic field  and for the powder sample, respectively.
The value of $J_c$ for $H||$($ab$) has ultra-high values of the order of $10^8$ A/cm$^2$, which is an order of magnitude larger than the value obtained when $H || c$. However, the powder sample has a $J_c$ which is at least three orders of magnitude lower than that of the single crystal (Table~SM1),
but similar to other polycrystalline iron-based superconductors \cite{Ma2012a,Shimoyama2014a}.
The low value of $J_c$ of the powder sample is not surprising because the magnetic behaviour is significantly affected by the extrinsic factors, such as grain morphology, surface roughness, inter-grain voids and interfaces \cite{Ma2012a}. These extrinsic factors could be reduced
by improving the synthesis process by adopting a ball-milling process, re-sintering and the optimization of
both the temperature and length of the preparation process.

  Fig.\ref{fig3}(d)-(e) shows that $J_c$  of CaKFe$_4$As$_4$ is robust and little affected by a magnetic field up to 16~T,
   reaching  a value of  $\sim 1.9(2) \times 10^8$~A/cm$^2$  for $H||$($ab$)    for 1~T and only dropping to $ \sim 0.7(2) \times 10^8$ ~A/cm$^2$ by 15~T.  These are some of the highest values obtained for any iron-based superconductor single crystal.
  This critical current value suggests strong vortex pinning, establishing this stoichiometric
   high-$T_c$ compound as a potential candidate for practical applications.
 To further emphasise this important aspect,
 we have plotted $J_c$ value for single crystal and the powder sample with respect to
 temperature for different magnetic fields  of 0 and 15~T, as shown in Figs.\ref{fig4}(a) and (b), respectively.
 We find that $J_c$ remains large over a large temperature range between 2 to 25~K, even with increasing magnetic field up to 15~T.
 Furthermore,  when comparing the temperature dependence of its critical current  (for $H || c$ at 0~T and 3~T)
  with other single crystals of iron-based superconductors  or
   conventional superconductors in Figure~\ref{fig4}(c)-(d), it is evident that
   CaKFe$_4$As$_4$  displays a ultra-high self-field $J_c$ around $10^7$~A/cm$^2$ for $H||c$ and these
  characteristics remain robust  up to 31~K in 3~T (up to a value of $10^6$~A/cm$^2$).
To further confirm these large $J_c$ values, we have measured additional crystals, as shown in Fig.\ref{fig3SM} and Table~SM1
  that show very similar values to those presented earlier;  a slight decrease in $J_c$ is noticed
      as the thickness of the sample increases from $2 \mu $m towards $15 \mu$m (see Table~SM1).
     These high $J_c$ values of  this stoichiometric clean superconductor
with a relatively low residual resistivity ($\rho_0 \sim 15.9 \mu \Omega$ cm) and close to an optimum doping
is surprising, as one would expect a high density of defects  to achieve effective pinning and large critical currents.

\begin{figure}[htbp]
	\centering
		\includegraphics[trim={0cm 0cm 20cm 0cm},width=0.95\linewidth,clip=true]{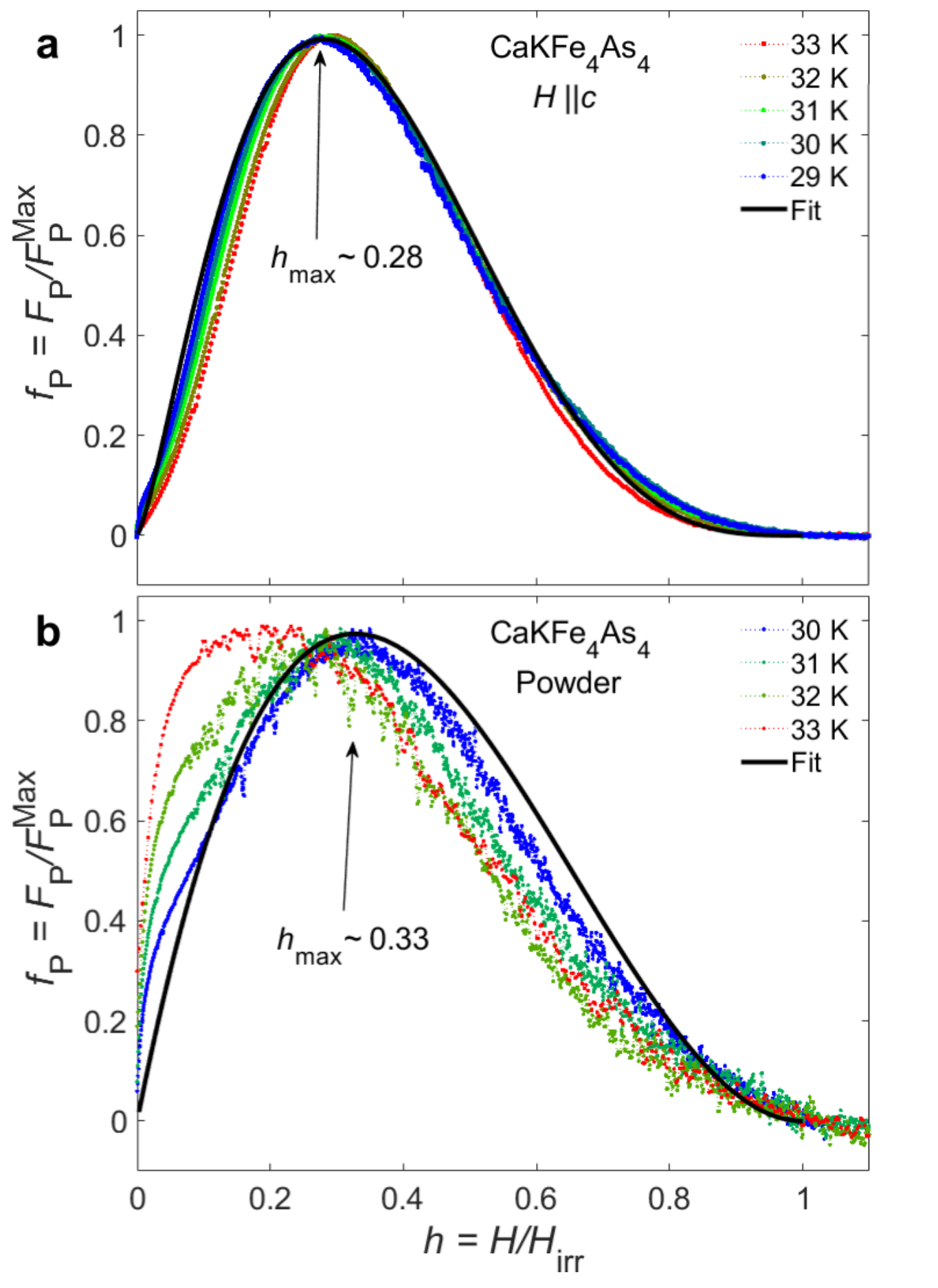}
	\caption{The scaling density of the normalized pinning force density, $f_P$, for different temperatures as a function of the reduced field, \textit{h}, based on the $f_P$(h) scaling function for (a) single crystal $H || c$ and (b) powder sample of CaKFe$_4$As$_4$. The solid line represents the fitted curve and data measured at different temperatures scaled together with the maximum reduced magnetic field at $h_{max} \sim 0.28$ for the single crystal in (a)
and   $h_{max} \sim 0.33$ for the powder in (b).}
	\label{fig5}
\end{figure}

In order to understand the nature of pinning, we study the temperature and field dependencies of vortex pinning force density, $F_P$,
 given by $F_P=\mu_{0} \textit{H} \times J_c$, when $\textit{H} \perp J_c$ .  According to the Dew-Hughes model \cite{Dew-Hughes1974}, if a dominant vortex pinning mechanism exists in certain temperature range, the normalized vortex pinning force density $f_P$ = $F_P$ / $F_P^{max}$ at different temperatures should collapse into one curve, as shown in Fig.\ref{fig5}.
The resulting scaling law  is given by $f_P$(\textit{h})=$h^p$$(1-h)^q$, where $F_P^{max}$ is the maximum pinning force density;
 the indices \textit{p} and \textit{q} are two parameters whose values depend on the origin of pinning mechanism and $h = H/H_{\rm irr}$ is the reduced field,
where the irreversibility field $\mu_{0} H_{\rm irr}$ is estimated as the magnetic field at which $J_c$($H$) reaches the background noise value, as defined in Fig.~\ref{fig3}a.

Figure~\ref{fig5}(a) and (b) shows the normalized pinning force density against the reduced magnetic field \textit{h}
of CaKFe$_4$As$_4$ estimated at different temperatures for the single crystal ($H || c$) and the powder sample.
As $H_{irr}$ exceeds the accessible magnetic field of 16~T at temperature below 27~K, only data closed to $T_c$ are presented ($T \geq 29$~K) .
The calculated maximum pinning force density ($F_P^{max}$) at 30~K
reaches a value of up to 9.58 GN/$\mathrm{m}^{3}$ for the single crystal
and a much reduced value of only 0.0007 GN/$\mathrm{m}^{3}$  for the  powder sample.
The normalized vortex pinning force density, $f_P$(h), collapse onto on a single curve with
 a $h_{max}\sim 0.28$ for the single crystal in Fig.~\ref{fig5}a,
 indicating that a  dominant pinning mechanism exists within this temperature range.
 For the powder sample (Fig.\ref{fig5}(b)), the peaks in $f_P$(h) are much broader,
  due to its granular nature, however the $h_{max}$ value is close to that of the single crystal.
 The fitting parameters to the $f_P$(\textit{h})
give \textit{p} = 1.4(1), \textit{q} = 3.7(3) and $h_{max} \sim 0.28$ for the single crystal $H || c$
and \textit{p} = 1.0(1), \textit{q} = 2.1(1) and $h_{max} \sim 0.33$ for the powder sample, respectively (Fig.\ref{fig5}).
 In the Dew-Hughes model for pinning mechanism, a value of the $h_{max} \sim 0.2$ corresponds to the surface pinning for normal center of core interaction, while $h_{max} \sim 0.33$ corresponds to the point core pinning, also known as small size normal pinning \cite{Dew-Hughes1974}.
 Thus, $h_{max}$ $\sim0.28-0.33$, observed for CaKFe$_4$As$_4$ suggests that vortex pinning
can be caused by a mixture of the surface and point core pinning of the normal centres,  similar to the Fe(Se,Te) systems \cite{Yadav2011}.

The obtained values of exponents \textit{p} and \textit{q} in CaKFe$_4$As$_4$
listed in Table~SM2 also support the idea that not a single model can describe the flux pinning mechanism \cite{Dew-Hughes1974} and
 the flux creep might have influence on the pinning force density.
However, similar values of \textit{p} and \textit{q} have been reported for other unconventional superconductors,
such as  FeTe$_{0.6}$Se$_{0.4}$ ($h_{max}$ = 0.28) and YBa$_2$Cu$_3$O$_7$ ($h_{max}$ = 0.33) \cite{Yadav2011,Klein1994,Wen1996}
(see Table~SM2).
In the 122 iron-based superconductors, the pinning mechanism depends on the doping level and dopant and $h_{max}$ tends to increase with increasing doping level
\cite{Ishida2017}. The highest $J_c$ is achieved  when $h_{max} \sim 0.40-0.45$ for the optimally-doped compounds \cite{Ishida2017}.
Single crystal of Ba$_{0.6}$K$_{0.4}$Fe$_2$As$_2$ shows only small-size normal pinning ($h_{max}$ = 0.33) \cite{Dew-Hughes1974},
whereas the values obtained for BaFe$_{1.8}$Co$_{0.2}$As$_2$ single crystal \cite{Ishida2017}
($h_{max}$ = 0.4, \textit{p} = 1.67, \textit{q} = 2)
are correlated to a dense vortex pinning nano-structure, likely due to the inhomogeneous distribution of cobalt ions.
A high $J_c$ is normally can be caused by a dense vortex pinning nano-structure mechanism.
Violation of the scaling behaviour in the underdoped and overdoped 122 iron-based superconductors  implies the existence of multiple pinning sources \cite{Ishida2017}. Stoichiometric CaKFe$_4$As$_4$
shows superconducting behaviour equivalent to an optimally-doped superconductor \cite{Ishida2017},
 but in the absence of inhomogeneities introduced by  chemical substitution.
Our scaling analysis implies that the proximity to
a optimally-doping regime together with
its mixed pinning mechanisms might be responsible for its high $J_c$.

\begin{figure}[htbp]
	\centering
	\includegraphics[trim={0cm 0cm 20cm 0cm},width=0.95\linewidth,clip=true]{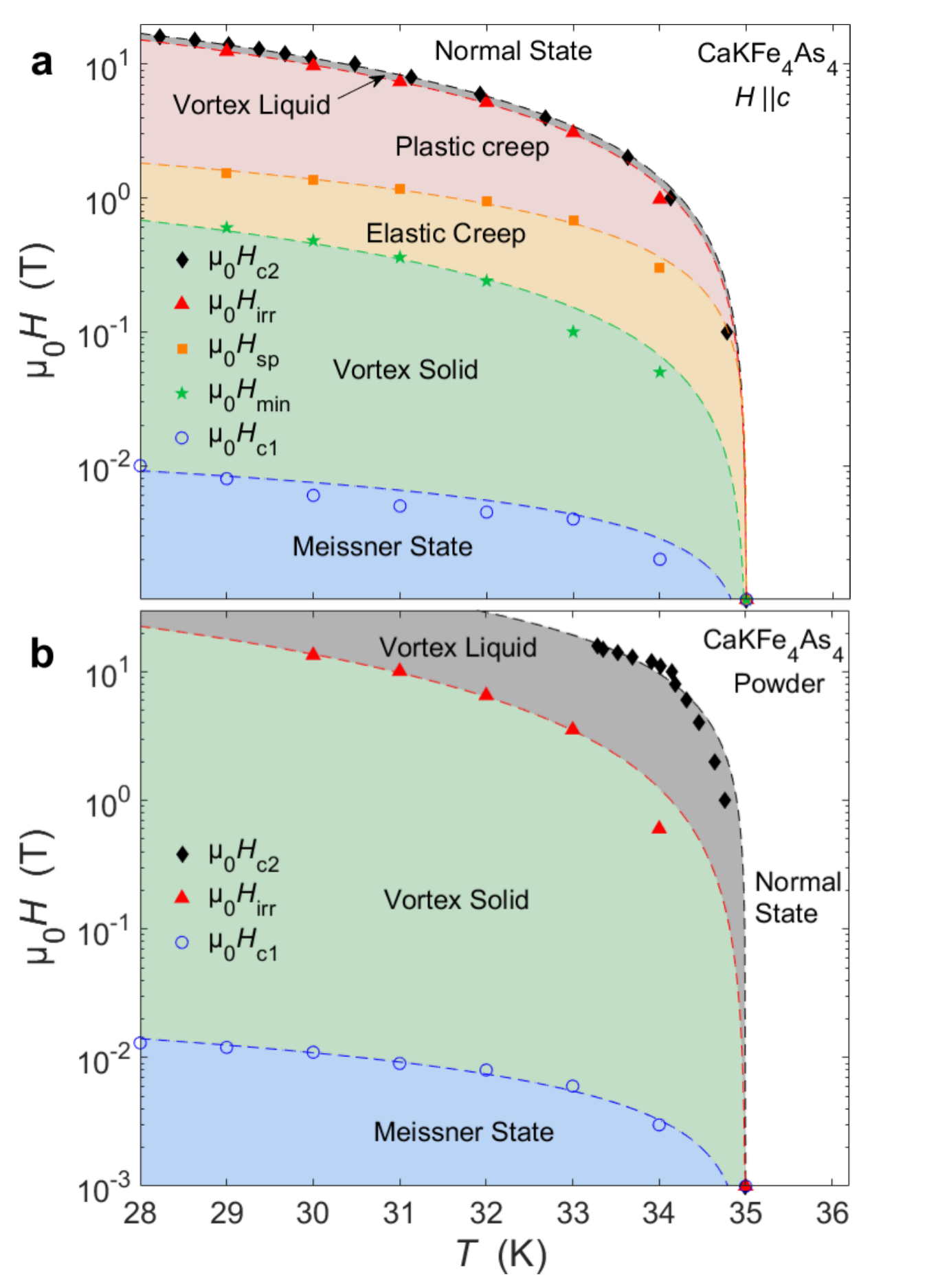}
		\caption{The vortex phase diagram of a) a single crystal when $H || c$ and b) polycrystalline sample of CaKFe$_4$As$_4$. The different magnetic fields separating different regions in the vortex phase
		diagram are defined in Fig.\ref{fig2} and 	Fig.\ref{fig3}. The dash lines are fits
		 using
		the empirical expression \textit{H}(\textit{T}) = \textit{H}(0)${(1 - T/{T}_{c})}^{n}$.}
		\label{fig6}
\end{figure}

Based on our magnetization studies, the vortex phase diagrams of  CaKFe$_4$As$_4$
for the single crystal (when $H||c$) and the powder sample are shown in Figs.~\ref{fig6}(a)-(b).
There are five different characteristic fields
separating different regions of the $H-T$ vortex phase diagram.
$H_{\rm min}$ and $H_{\rm sp}$ are the magnetic field located at the
valley point  and the second magnetization peak, respectively,
in the $J_c$ versus \textit{H} plot,
as defined in Fig.\ref{fig3}(d).
The irreversibility field $H_{irr}$ is defined
 in Fig.\ref{fig3}(d)
and $H_{c1}$ and $H_{c2}$ have been discussed previously in detail in Fig.~\ref{fig2}.
In the powder sample, there is no second magnetization peak.
The irreversibility line, $H_{\rm irr}$,  is very close to $H_{c2}$ in single crystals
(Fig.\ref{fig6}(a)) and this is a very important feature for CaKFe$_4$As$_4$ to be technologically useful,
as the irreversibility line demarcates the field at which vortex flow is unpinned and magnetic irreversibility sets in.
On the other hand,  the powder sample has  $H_{\rm c2}$  a factor 4 larger than
in single crystals and vortex liquid state is  further extended, as shown in Fig.\ref{fig6}(b).
 This suggests that grain boundaries increase the density of the pinning centers to give rise to
 larger upper critical fields, whereas the poor grain connectivity in powder is detrimental to supporting large critical currents, as
 they are significantly reduced.

 Different regimes of vortex dynamics \cite{Ni2008a,Salem-Sugui2010}, marked elastic creep and plastic creep, separated by the $H_{\rm sp}$ line, are shown in the phase diagram in Fig.\ref{fig6}a. In general, the vortex phase diagram is determined by the competition between the elastic energy of the vortex lattice, the pinning energy and  thermal energy \cite{Ishida2017}. The second magnetization peak effect is commonly associated with an order-disorder transition, which occurs at $H_{min}$ when the pinning energy exceeds the elastic energy \cite{Brandt1999}. The motion of the vortices is governed by the elastic force in the elastic and plastic creep regions, and $H_{sp}$ corresponds to the threshold field of the elastic-plastic crossover.
  In the plastic creep region, as the magnetic field increases
 the vortex lattice softens and vortices are pinned more easily, resulting in an increase in $J_c$.
 As the magnetic field increases
  the pinning force and thermal energy affect the vortices  significantly and
 the vortices can easily move from one pinning center to another in the vortex liquid state, which is above $H_{irr}$.
 In the case in which the pinning energy does not exceed the elastic energy in the entire field range, the SMP disappears, owing to the absence of an order-disorder transition, as in the case of powder sample in Fig.~\ref{fig6}(b).

The vortex phase diagram of CaKFe$_4$As$_4$ single crystal is very similar
to that reported for the (Ba,K)Fe$_2$As$_2$ single crystal \cite{Yang2008},
with $H_{\rm min}$ having almost the same value
but $H_{\rm sp}$ and $H_{\rm irr}$ being higher in the 122 compounds.
 The temperature dependence of the different characteristic fields in Fig.~\ref{fig6}(a) can be fitted using
 the empirical expression \textit{H}(\textit{T}) = \textit{H}(0)${(1 - T/{T}_{c})}^{n}$, with  \textit{n} varying between
 1.3(1) and 0.6(1).
 The obtained $n$ value corresponding to the $H_{min}$ curve is $n$=1.2(1),  similar to values reported for other iron-based superconductors \cite{Das2011,Pramanik2011a}, whereas the $H_{sp}$  and its corresponding exponent, $n$=0.83(6),  are much smaller than those found in other systems \cite{Yang2008,Das2011,Pramanik2011a}.
  Furthermore, the vortex phase diagram of CaKFe$_4$As$_4$,  delimited by the temperature dependence of  $H_{irr}$,
  shows a high exponent value of $n$=1.3(1),   which cannot be understood within the framework of conventional superconductivity.

 CaKFe$_4$As$_4$ is a stoichiometric rather clean superconductor exhibiting very high critical current values,
 rather unexpected, as a high density of defects is necessary to achieve such high values of $J_c$.
 However, recent STM studies find a disordered vortex lattice up to 8~T, with the vortices  pinned at
the locations where the superconducting order parameter is strongly suppressed due to pair breaking and
 the  vortex  core  size decreases with increasing field which will give rise to very large upper critical fields  \cite{Fente2017}.
The low-temperature vortex core size determined by STM is $\sim 1.3$~nm, in reasonable
agreement with the value estimated from upper critical field study of $\sim 1.9$~nm, demonstrating the
importance of point pinning in this material.
Furthermore, the surface topography measured by STM exhibits steps on the surface due to structural defects, surface reconstruction and small stripes, which as possible sources of surface pinning. Other types of two-dimensional surface pinning may be influenced by the anisotropic layered structure and stacking faults that may occur between Fe layers along the $c$-axis.

The observed high $J_c$ values are likely to be generated by intrinsic effects
due to the ionic size  variation that may occur in the crystallographic structure and  the two-dimensional surface defects in
 CaKFe$_4$As$_4$. This structure has relevant differences compared to a 122 system,
 due to the absence of a glide plane and the positions of the Fe-As layers generating different Fe-As distances.
 The small structural defects can cause mean free path fluctuation induced pinning,
 which is an effective way to enhance $J_c$ \cite{Griessen1994}.
 Furthermore, close to the optimal doping, as in the case of  CaKFe$_4$As$_4$,  the vortex core energy of the flux lines can be enhanced  \cite{Putzke2014}.
Thus, the high vortex core energy and the strength of the intrinsic depairing effect can be
key factors responsible for the extremely high $J_c$ value in this compound.
Generally, the intrinsically  high depairing critical density is significantly reduced due to the presence
of extrinsic effects, such as weak links and grain mismatch, as for our powder sample of CaKFe$_4$As$_4$.

\section{Conclusions}

In summary, we have investigated the
critical current densities and the vortex phase diagram of an optimally-doped stochiometric superconductor, CaKFe$_4$As$_4$,
both in single crystal and powder form.
Interestingly, we find that this material exhibits extremely large   $J_{c}$ values
up to $10^{8}$ A/$\mathrm{cm}^{2}$ at low temperatures,
which are some of the highest values for a iron-based superconductor.
The critical current density for both in-plane and out-plane orientations
 has a very weak temperature dependence in the temperature range up to 25~K.
The magnetization curve shows very large hysteresis loops,  suggesting strong flux pinning.
We detect the fish-tail effect when $H || c$, similar to other optimally-doped 122 superconducting
 single crystals, but this effect is missing for the $H||$($ab$) orientation and the powder sample.
The flux pinning force density
suggest the existence of surface and the point core pinning of  the vortices in low field regime.
The extremely-high critical current density of  CaKFe$_4$As$_4$,
suggests that the vortex core energy of the flux lines can be enhanced in
this optimally doped superconductor and the reduced symmetry of As-Fe bonds
could play an important role in pinning.
 Furthermore, in the powder sample
 the critical current densities are
 significantly reduced  due to the  weak grain connectivity,
  as often found in other iron-based superconductors.
The ultra-high critical current densities
 and the vortex phase diagram of CaKFe$_4$As$_4$
place this stoichiometric high-$T_c$  superconducting family
as a realistic contender for practical applications.

\section{Acknowledgements}
We thank A. Iyo, Y. Yoshida and H. Eisaki for the provision of the polycrystalline sample used in this study and useful discussions.
The research was funded by the Oxford Centre for Applied Superconductivity (CFAS) at Oxford University.
We also acknowledge financial support of the John Fell Fund of the Oxford University.
Work done at Ames Laboratory was supported by the U.S. Department of Energy, Office of Basic Energy Science, Division of Materials Sciences and Engineering. Ames Laboratory is operated for the U.S. Department of Energy by Iowa State University under Contract No. DE-AC02-07CH11358.
W. M. was funded by the Gordon and Betty Moore Foundation’s EPiQS Initiative through Grant GBMF4411.
 AIC acknowledges an EPSRC Career Acceleration Fellowship (EP/I004475/1).

\bibliography{Ca1144_bib_may2018}

\setcounter{figure}{0}
\newcommand{\blue}{\textcolor{blue}}
\newcommand{\bdm}[1]{\mbox{\boldmath $#1$}}
\renewcommand{\thefigure}{SM\arabic{figure}} % changes FIG.~1 to FIG.~SM1
\renewcommand{\thetable}{SM\arabic{table}} % changes FIG.~1 to FIG.~SM1

\begin{figure*}[htbp]
	\centering
	\includegraphics[trim={0cm 0cm 0cm 0cm},width=1\linewidth,clip=true]{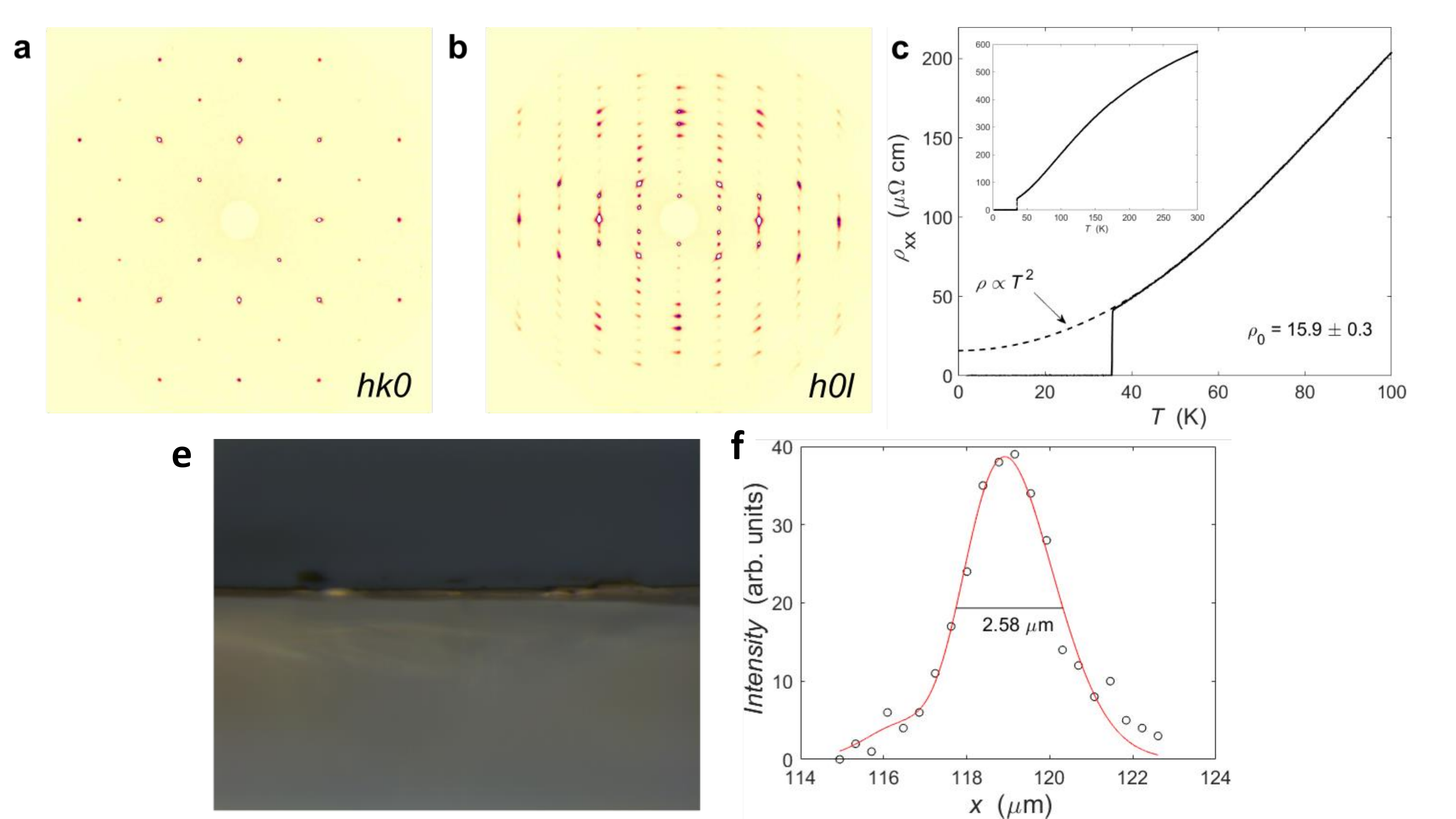}
	\caption{X-ray diffraction pattern of a) the ($hk0$) and b) the ($h0l$) plane for a single crystal from the same batch like the measured crystal.
		c) Resistivity data showing the very high quality of our single crystals giving
		an extrapolated low-temperature residual resistivity of $\rho_0 \sim 15.9(3) \mu \Omega $ cm.
The residual resistivity ratio, RRR ( defined as $\rho_{300K}$ / $\rho_{36K}$) is $\sim 14$,
in agreement with previous reports \cite{Meier2016},
and together with the small superconducting transition width ($\Delta$$T_c$) of 0.3 K,
reflect the very high-quality of our single crystal. e) Optical  image of the single crystal used in the magnetization studies.
f) The estimation of the sample thickness from e) by fitting to a gaussian distribution of the intensity. }
	\label{fig1SM}
\end{figure*}

\begin{figure*}[htbp]
	\centering
	\includegraphics[trim={0cm 0cm 0cm 0cm},width=0.8\linewidth,clip=true]{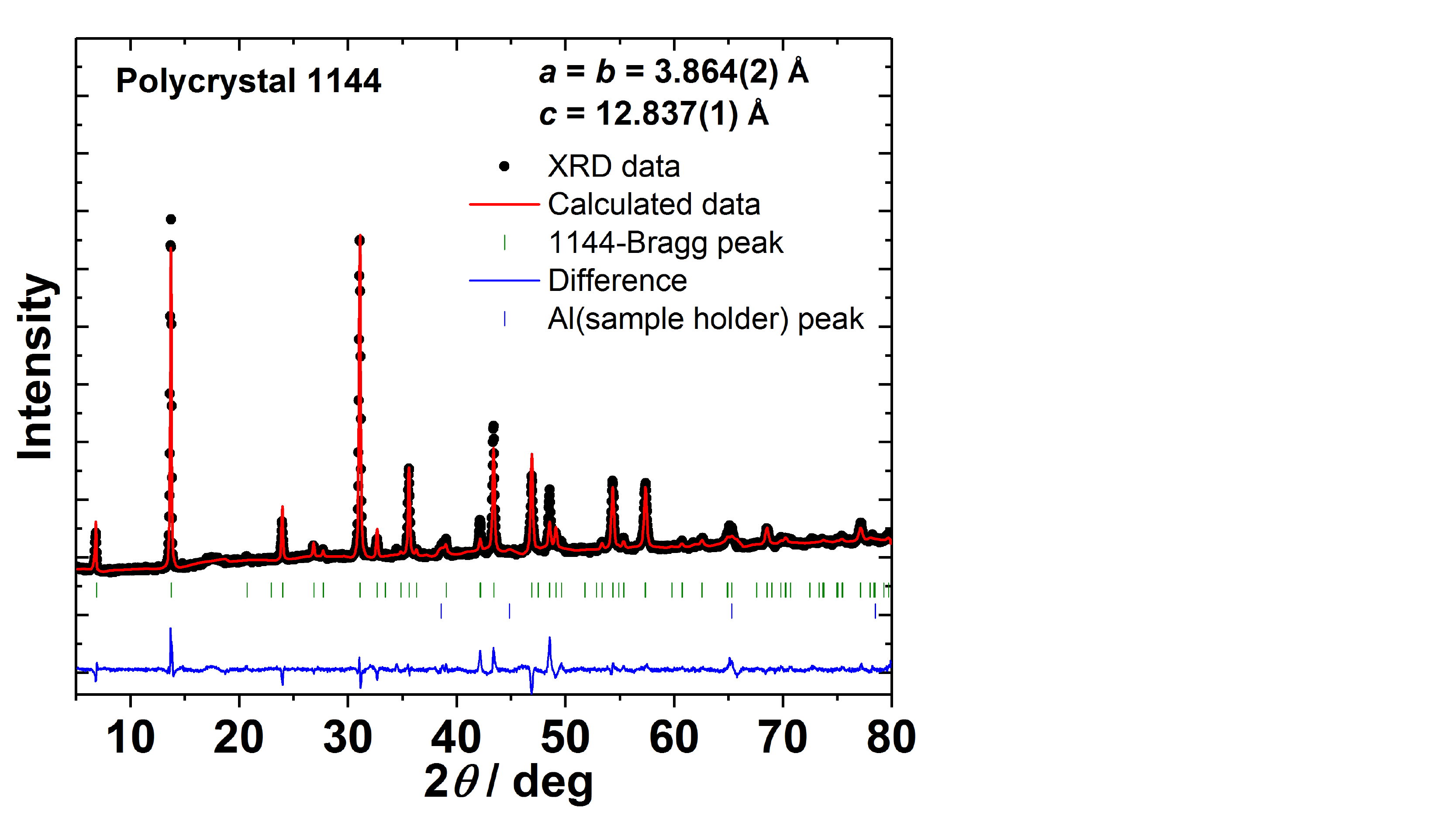}
		\caption{Powder x-ray diffraction patterns (XRD) and Rietveld refinement of the polycrystalline sample of CaKFe$_4$As$_4$ used in this study.
The X-ray data show that this is a single phase compound  with the extracted lattice parameters closed to those reported for the single crystal
 ($a$ = 3.8659 \AA, $c$ = 12.884 \AA \cite{Meier2016}).}
	\label{fig2SM}
\end{figure*}

\begin{table*}[t]
\label{Table_SM1}
	\centering
	\setlength\tabcolsep{0.00000001\linewidth}
	\caption{The transition temperature, $T_{c}$,
the slope  of the upper critical field close to $T_c$,
upper critical field, $H_{c2}$, and the lower critical field, $H_{\rm c1}^*$, and the critical current density, $J_c$, extrapolated at zero field,
 of our single crystal (S1), an additional crystal S4 (thickness of $15~\mu$m) and the powder sample of CaKFe$_4$As$_4$.}
	\begin{center}		
		\begin{tabular}{ p{0.2\linewidth} p{0.12\linewidth} p{0.12\linewidth} p{0.12\linewidth} p{0.12\linewidth} p{0.13\linewidth} p{0.08\linewidth} p{0.12\linewidth} p{0.12\linewidth} }
			\hline
\hline
			\textbf{Parameters} & \textbf{Single Crystal} & \textbf{Single crystal} &	\textbf{Polycrystalline}\\
			& $H||$($ab$) & $H||c$ & \\
			\hline
			      $T_c$(K)  &	35.0(1)	 & 35.1(2) 	& 34.9(2) \\
					  $d H_{c2}/dT$(T/K)                & 5.7  & 2.6  &  	10.1         \\
					  $H_{c2}$(0)   &   135  & 62  & 241 \\
					  $H^*_{c1}$(2~K)(T) &	0.1114 & 0.0220 &	0.0549 \\
						$J_c$(A/$\mathrm{cm}^2$) (5~K)(S1) & 1.9(2)$\times$$10^8$ & 0.9(2)$\times$$10^7$ & 4.4$\times$$10^3$ \\
			$J_c$(A/$\mathrm{cm}^2$) (5~K) (S2) & 0.8(2)$\times$$10^8$ & 0.5(2)$\times$$10^7$ & -- \\
$J_c$(A/$\mathrm{cm}^2$) (5~K) (S4) & -- & 0.4(2)$\times$$10^7$ & -- \\
			\hline
\hline
		\end{tabular}
	\end{center}
\end{table*}

\begin{figure*}[htbp]
	\centering
\includegraphics[trim={0cm 0cm 0cm 0cm},width=1\linewidth,clip=true]{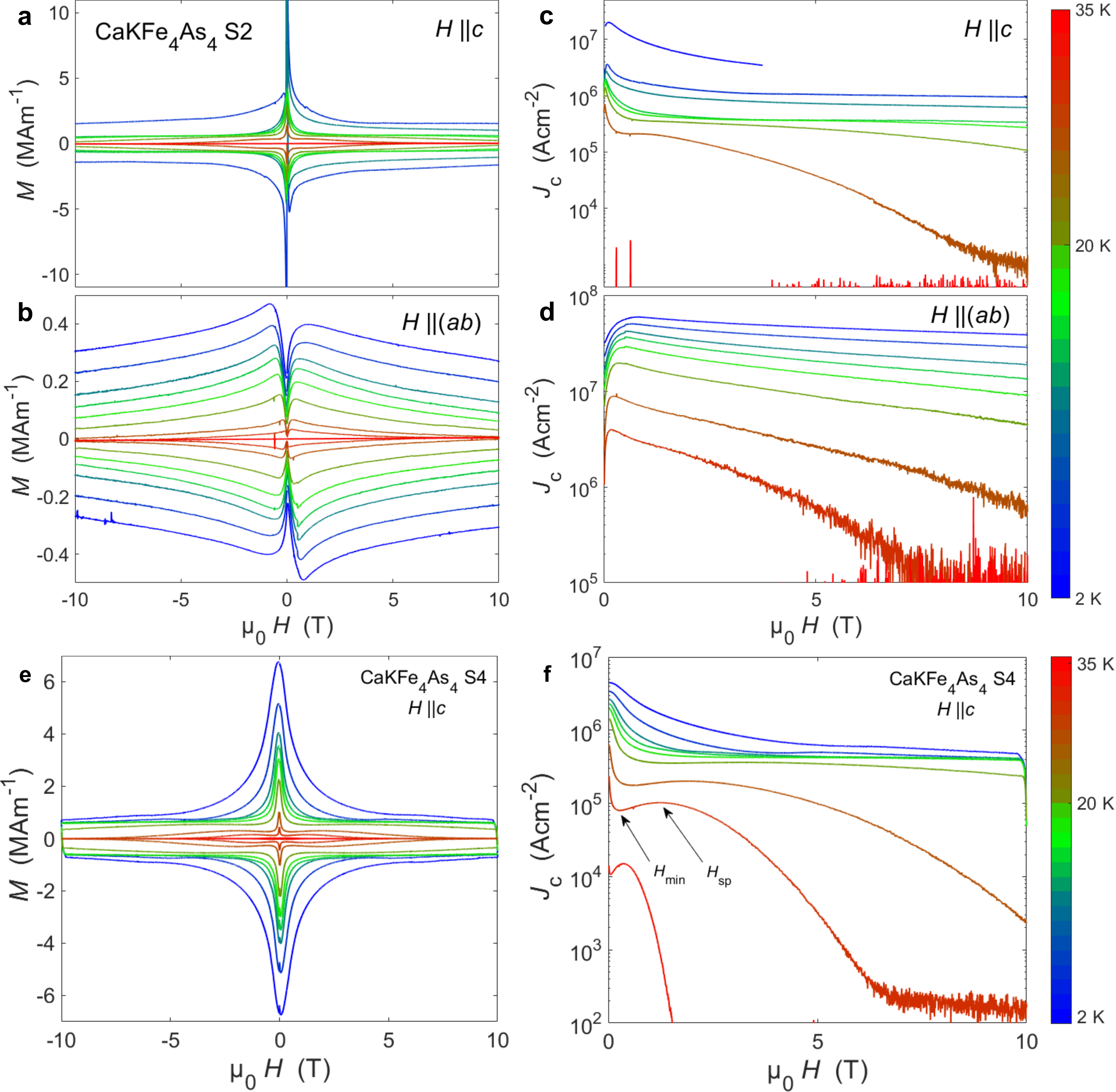}
	\caption{ a), b) Magnetization and c), d) critical current density for
a different single crystals,  S2 (with thickness around 12~$\mu$m) measured with $H||c$ and $H||$($ab$), respectively.
 e) Magnetization and f) critical current density for the single crystal S4 (with thickness around 15~$\mu$m) measured with $H||c$.}
	\label{fig3SM}
\end{figure*}

\begin{table*}[t]
	\centering
	\setlength\tabcolsep{0.00000001\linewidth}
	\caption{Summary of the pinning force scaling parameters \textit{p}, \textit{q} and $h_{max}$, for different superconductors.}
		\begin{center}		
		\begin{tabular}{ p{0.25\linewidth} p{0.1\linewidth} p{0.1\linewidth} p{0.22\linewidth} p{0.22\linewidth} }
			\hline
			\textbf{Single crystal} & \textbf{\textit{p}} & \textbf{\textit{q}} &	\textbf{$h_{max}$ = \textit{p}/(\textit{p}+\textit{q})} &	\textbf{Peak position} \\
			\hline
			CaKFe$_4$As$_4$ &	1.4(1)	& 3.7(3)	& 0.28 &	0.28 \\
			\hline
			$\mathrm{Ba}_{0.6}$$\mathrm{K}_{0.4}$$\mathrm{Fe}_2$$\mathrm{As}_2$ \cite{Yang2008} &	1 & 2 &	0.33 &	0.33 \\
			\hline
			$\mathrm{Ba}_{0.68}$$\mathrm{K}_{0.32}$$\mathrm{Fe}_2$$\mathrm{As}_2$ \cite{Sun2009} &	 &  &	 &	0.43 \\
			\hline
			Ba$\mathrm{Fe}_{1.8}$$\mathrm{Co}_{0.2}$$\mathrm{As}_2$ \cite{Yamamoto2009} &	1.67 & 2 &	0.45 &	0.45 \\
			\hline
			Ba$\mathrm{Fe}_{1.85}$$\mathrm{Co}_{0.15}$$\mathrm{As}_2$ \cite{Sun2009} &	 &  &	 &	0.37 \\
			\hline
			Ba$\mathrm{Fe}_{1.29}$$\mathrm{Ru}_{0.71}$$\mathrm{As}_2$ \cite{Sharma2013} &	1.95	& 2.5 &	0.44 &	0.45 \\
			\hline
			Ba$\mathrm{Fe}_{1.91}$$\mathrm{Ni}_{0.09}$$\mathrm{As}_2$ \cite{Sun2009} &		&  &	 &	0.32 \\
			\hline
			Fe$\mathrm{Te}_{0.7}$$\mathrm{Se}_{0.3}$ &	 &  &	 &	0.27 \\
			\hline
			Fe$\mathrm{Te}_{0.6}$$\mathrm{Se}_{0.4}$ \cite{Yadav2011} &	1.54 & 3.8 &	0.28 &	0.28 \\
			\hline
			$\mathrm{K}_x$$\mathrm{Fe}_{2-y}$$\mathrm{Se}_2$ \cite{Lei2011}&	0.86 & 1.83 &	0.32 &	0.33 \\
			\hline
			Y$\mathrm{Ba}_2$$\mathrm{Cu}_3$$\mathrm{O}_7$ \cite{Klein1994} &	2 & 4 &	0.33 &	0.33 \\
			\hline
			Nd$\mathrm{Ba}_2$$\mathrm{Cu}_3$$\mathrm{O}_{7-d}$ \cite{Koblischka1999}&	1.48 & 2.23 &	0.45 &	0.45 \\
			\hline
			$\mathrm{Sm}_{0.5}$$\mathrm{Eu}_{0.5}$$\mathrm{Ba}_2$$\mathrm{Cu}_3$$\mathrm{O}_{7}$ \cite{Wen1996}&	2.08 & 3.56 &	0.37 &	0.37 \\
			\hline
		\end{tabular}
	\end{center}
\end{table*}

\end{document}